\documentstyle[11pt,newpasp,twoside,epsf]{article}
\def\edcomment#1{\iffalse\marginpar{\raggedright\sl#1\/}\else\relax\fi}
\reversemarginpar

\begin{document}
\title{Searching for Faint Companions with the TRIDENT Differential Simultaneous Imaging Camera}
  \author{Christian Marois, Daniel Nadeau,  Ren\'{e} Doyon, Ren\'{e} Racine}
\affil{D\'{e}partement de physique, Universit\'{e} de Montr\'{e}al, C.P. 6128, Succ. A,\\ Montr\'{e}al, QC, Canada H3C 3J7}
\author{Gordon A.~H. Walker}
\affil{1234 Hewlett Place, Victoria, BC, Canada~V8S~4P7}

\begin{abstract}
We present the first results obtained at CFHT with the TRIDENT infrared camera, dedicated to the detection of faint companions close to bright nearby stars.  Its main feature is the acquisition of three simultaneous images in three wavelengths (simultaneous differential imaging) across the methane absorption bandhead at $1.6\mu$m, that enables a precise subtraction of the primary star PSF while keeping the companion signal. Gl229 and 55Cnc observations are presented to demonstrate TRIDENT subtraction performances. It is shown that a faint companion with a $\Delta H$ of 10 magnitudes would be detected at $0.5\arcsec$ from the primary.
\end{abstract}

\section{Introduction}
In the past few years, our group has developed the specialized infrared camera TRIDENT (Marois et al. 2000a, 2002) to search for faint companions (brown dwarfs and Jovian planets) at a projected distance of 5AU to 50AU from stars in the solar neighborhood.  This corresponds to the range of orbital distance for the massive planets of the Solar system. It is therefore of special interest to test the uniqueness of our planetary system. Ground detection of faint companions is difficult because of the atmospheric turbulence that distorts the star PSF and optical aberrations that produce quasi-static PSF structures. The main feature of TRIDENT is to acquire three simultaneous images in three distinct narrow spectral bands, making possible very good atmospheric speckle and optical aberration calibration to subtract the stellar PSF. If the three simultaneous wavelengths are carefully chosen, it is possible to enhance the star/companion contrast after image combinations by selecting special spectral features that are typical of the companion and not of the star (Smith 1987; Rosenthal, Gurwell, \& Ho 1996; Racine et al. 1999; Marois et al. 2000b). In TRIDENT, the three wavelengths (1.567~$\mu$m, 1.625~$\mu$m and  1.680~$\mu$m, 1\% bandwidth) have been selected across the 1.6 $\mu$m methane absorption bandhead that is only present in the spectrum of cold ($T_{\rm{eff}}$ $<$ 1470~K, Fegley and Lodders (1996)) substellar objects. This paper describes the first results obtained with TRIDENT.

\section{Observations}
Data were obtained on 2001 July 8-12 and on 2001 November 21-24, at the f/20 focus of the 3.6m CFHT adaptive optics bonnette PUEO (Rigaut et al. 1998) with TRIDENT. Flat fields and darks were obtained during the day. Three types of observing techniques were used. In July, data well inside the linear regime of the detector were acquired to test the PSF stability. Total integration time was typically 1h per target. Seeing conditions for this run were medium to good (AO strehl of 0.5 in $H$ band). In November, reference stars were acquired and instrument rotations ($+/-$ 90 degrees by steps of 2 degrees) were done to calibrate or smooth non-common path aberrations discovered to be the limiting factor in the July mission data. Saturated and non-saturated images were acquired to minimize readout noise. Total integration time was 1h30 with medium seeing conditions (AO strehl from 0.15 to 0.4 in the $H$ band). In total, 35 stars were observed during these two missions with spectral type ranging from B to M. Some stars were observed on different nights to acquire more photons and study PSF stability. See Marois et al. 2003 for more details about data reduction and analysis. Figure 1 shows the Gl229 system with the star PSF removed.  

\begin{figure}[h]
\plotfiddle{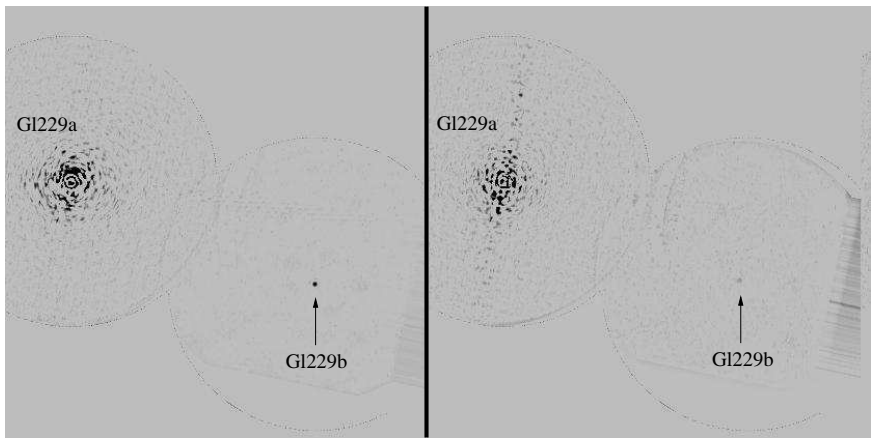}{5cm}{0}{43}{43}{-162}{-10}
\caption{The Gl229 system with the star PSF removed (15 minutes integration). The left frame shows the difference between the $1.567~\mu$m and the $1.680~\mu$m images. In the right frame, the difference between the $1.625~\mu$m and the $1.680~\mu$m images is presented. Gl229 methane brown dwarf companion (Gl229b) is visible in the lower right corner of the left frame.}
\end{figure}
If a companion contains methane in its atmosphere, like Gl229b, it will be bright in the $1.567~\mu$m filter and faint in the other two filters. The subtraction of the $1.567~\mu$m and the $1.680~\mu$m images thus shows a bright companion while the subtraction of the $1.625~\mu$m and the the $1.680~\mu$m images does not. The companion magnitude through the three narrow band filters is thus used to confirm the presence of methane in the object atmosphere and the substellar nature of the object. A spectral class can be derived by estimating the companion contrast ratios between the three filters and then fit them with the Burgasser et al. 2002 T-type spectral classification. A proper motion follow-up would be necessary to confirm the gravitationnal bond between the object and the star.
\clearpage
TRIDENT can also detect companions with no methane in their atmosphere. Since the three PSFs need to be scaled to properly subtract the PSF structure, the companion-star separation is thus evolving with wavelength. Figure 2 shows the 55Cnc star PSF subtraction residuals with a reference star calibration and a simulated L dwarf (no methane) companion as seen at CFHT and scaled to an 8~m telescope. In this situation, the companion is equally bright (no methane) in the three filters, so it is not possible to confirm directly the substellar nature of the object. Two observations a few months apart and photometric analysis would be necessary to confirm companionship and the substellar nature of the companion. Figure 3 shows the detection threshold for the differential simultaneous imaging technique. It can be seen that the PSF subtraction residuals are 3 magnitudes brigther than the photon noise limit at $0.5\arcsec$. PSF evolution due to optical flexions and atmospheric variations is the main limitation when attempting high precision PSF calibrations and subtractions.

\begin{figure}[h]
\plotfiddle{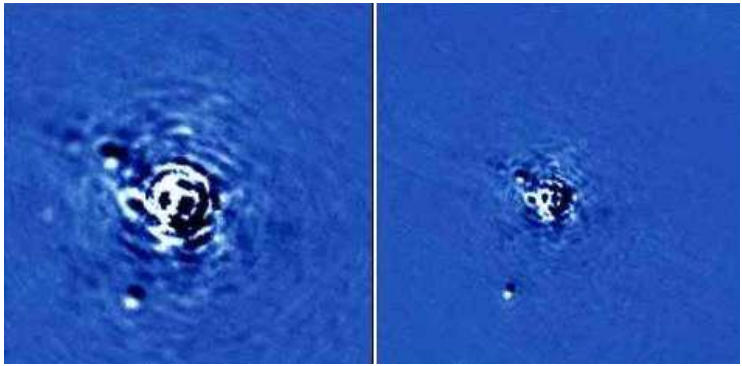}{3.8cm}{0}{45}{45}{-138}{-10}
\caption{The 55Cnc system with the star PSF removed (1h integration). A reference star was used to further enhance the detection threshold by subtracting aberrations in each optical path. A simulated L dwarf companion ($0.8\arcsec$ away and 8 magnitudes fainter) was added. The left frame shows the difference between the $1.567~\mu$m and the $1.680~\mu$m 55Cnc images as seen with the TRIDENT camera at CFHT. The right frame shows the same residuals but scaled to an 8~m telescope.}
\end{figure}
\section{Conclusion - TRIDENT new developments}
Typical attenuation in good seeing conditions is 10 magnitudes in the $H$ band at $0.5\arcsec$, as good as or better than what has been done on any other telescope. Reference star offered good but not perfect PSF calibration, mainly because of the PSF structure slow evolution with telescope pointing and atmospheric variations. Photon noise limited subtraction requires the minimization of optical aberrations. To increase PSF correlation, and thus subtraction performances, a new optical design is presently under evaluation that mainly consists of a better beam splitter and filters located near the detector to minimize wavefront degradation. The goal is photon noise limited performance for 1h integrations.

This work is supported in part through grants from NSERC, Canada and from Fonds FQRNT, Qu\'{e}bec.
\begin{figure}[h]
\plotfiddle{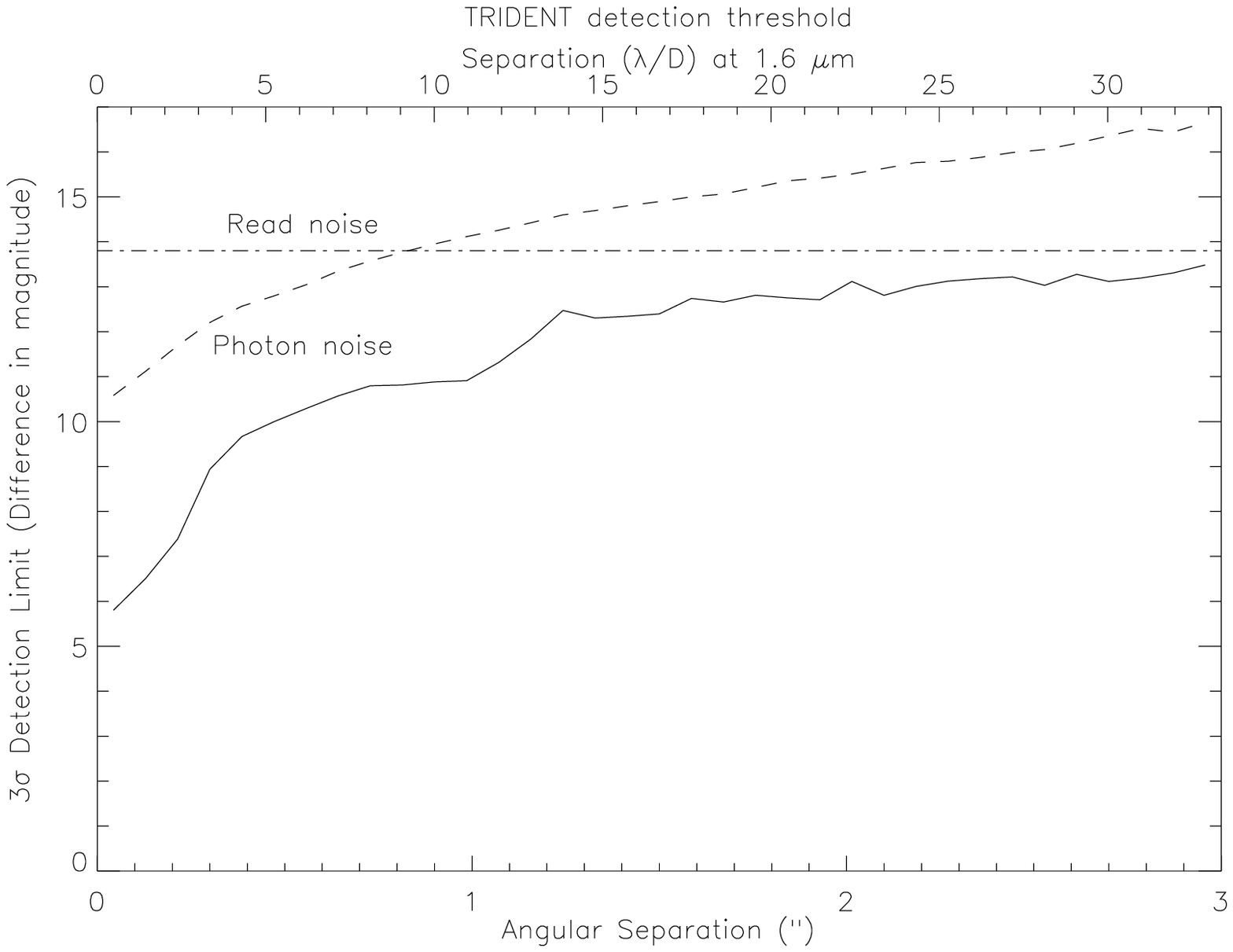}{5.5cm}{0}{50}{50}{-150}{-190}
\caption{TRIDENT detection threshold (solid line) for differential simultaneous imaging with reference star subtraction from the November 2001 CFHT $\upsilon$ And (object) and $\chi$ And (reference) data sets (1h integration per target). The dash and dash-dot lines correspond respectively to the estimated 3$\sigma$ photon noise and read noise limits.}
\end{figure}


\begin{thebibliography}
\bibitem[Bugasser et al. (2002)]{Burgass2002} Burgasser, A.~J., Kirkpatrick, J.~D., Brown, M.~E., Reid, I.~N., Burrows, A., Liebert, J., Matthews, K., Gizis, J.~E., Dahn, C.~C., Monet, D.~G., Cutri, R.~M., Skrutskie, M.~F. 2002, \apj, 564, 421
\bibitem[Fegley & Lodders (1996)]{Feg96} Fegley, B.~J.,  Lodders, K. 1996, \apjl, 472, L37
\bibitem[Marois et al. (2000a)]{Marois2000a} Marois, C., Doyon, R., Racine, R., Nadeau, D. 2000a SPIE, 4008, 788
\bibitem[Marois et al. (2000b)]{Marois2000b} Marois, C., Doyon, R., Racine, R., Nadeau, D. 2000b \pasp, 112, 91
\bibitem[Marois et al. (2002)]{Marois2002a} Marois, C., Nadeau, D., Doyon, R., Racine, R., Riopel, M., Vall\'{e}e, P. 2002, SPIE, 4860, in press
\bibitem[Marois et al. (2003)]{Marois2003} Marois, C., Nadeau, D., Doyon, R., Racine, R., Walker, G.~A.~H. 2003, IAU, 211, in press
\bibitem[Smith (1987)]{Smith87} Smith W.~H. 1987, \pasp, 99, 1344
\bibitem[Racine et al.(1999)]{Racine99} Racine, R., Walker, G.~A.~H., Nadeau, D., Doyon, R., Marois, C. 1999, \pasp, 111, 587
\bibitem[Rigaut et al. (1998)]{Rigaut98} Rigaut, F., Salmon, D., Arsenault, R., Thomas, J., Lai, O., Rouan, D., V{\' e}ran, J.~P., Gigan, P., Crampton, D., Fletcher, J.~M., Stilburn, J., Boyer, C., Jagourel, P. 1998, \pasp, 110,152
\bibitem[Rosenthal et al. (1987)]{Rosen87} Rosenthal, E.~D., Gurwell, M.~A., Ho, P.~T.~P. 1996, Nature, 384, 243
\end{thebibliography}
\end{document}